\documentclass[aps,nofootinbib]{revtex4-1}
\pdfoutput=1
\usepackage{graphicx,subfigure}
\usepackage{amsmath,amssymb}
\usepackage{color}
\usepackage{mathrsfs}  
\usepackage{ulem}

\def\beq{\begin{equation}}
\def\eeq{\end{equation}}
\def\beqr{\begin{eqnarray}}
\def\eeqr{\end{eqnarray}}
\def\bdpm{\begin{displaymath}}
\def\edpm{\end{displaymath}}
\def\half{\frac{1}{2}}
\definecolor{lgray}{gray}{0.6}

\newcommand{\tbt}{t_\beta}
\newcommand{\tth}{t_\theta}
\newcommand{\cth}{c_\theta}
\newcommand{\sth}{s_\theta}
\newcommand{\ai}{\textrm{i}}
\newcommand{\msbar}{\overline{\textrm{MS}}}
\newcommand{\Veff}{V_\textrm{eff}}

\begin{document}

\title{
Conformal invariance and singlet fermionic dark matter
}

\author{Yeong Gyun Kim}
\email{ygkim@gnue.ac.kr}

\affiliation{ 
Department of Science Education, 
Gwangju National University of Education, Gwangju 61204, Korea
}

\author{Kang Young Lee}
\email{kylee.phys@gnu.ac.kr}

\affiliation{ 
Department of Physics Education and
Research Institute of Natural Science,
Gyeongsang National University, Jinju 52828, Korea
}

\author{Soo-hyeon Nam}
\email{glvnsh@gmail.com}

\affiliation{ 
Department of Physics, 
Korea University, Seoul 02841, Korea
}

\date{\today}

\begin{abstract}

We study a classically scale-invariant model with an electroweak singlet complex scalar mediator together
with an anomaly free set of two fermionic dark matters.
We introduce $U(1)_X$ gauge symmetry with a new charge $X$ in the dark sector
in order to stabilize the mass of the scalar singlet with a new gauge boson. 
Our conformally invariant scalar potential generates the electroweak symmetry breaking 
via the Coleman-Weinberg mechanism, 
and the new scalar singlet acquires its mass through radiative corrections of 
the fermionic dark matters and the new gauge boson as well as of the SM particles.
Taking into account the collider bounds, we present the allowed
region of new physics parameters satisfying the recent measurement of relic abundance. With the
obtained parameter sets, we predict the elastic scattering cross section of the new singlet fermions
into target nuclei for a direct detection of the dark matter. 
We also discuss the collider signatures and future discovery potentials
of the new scalar and gauge boson.

\end{abstract}

%\pacs{PACS numbers:12.60.Fr,12.60.Cn,14.80.Cp}
\maketitle

%%%%%%%%%%%%%%%%%%%%%%%%%%%%%%%%%%%%%%%%%%%%%%%%%%%%%%%%%%%%%%%%%
\section{Introduction}
%%%%%%%%%%%%%%%%%%%%%%%%%%%%%%%%%%%%%%%%%%%%%%%%%%%%%%%%%%%%%%%%%

The discovery of a Higgs-like boson with $\sim$ 125 GeV mass at the Large Hadron Collider (LHC) completes
the Standard Model (SM) particle spectrum \cite{Aad:2012tfa, Chatrchyan:2012xdj}. 
However, the existence of such a (seemingly) fundamental lightish scalar particle raises a long-standing problem. 
The so-called naturalness problem essentially states 
that the Higgs mass parameter seems unnaturally small compared to the Planck scale
at which the SM or physics beyond the SM at the electroweak scale is  unified with the gravitational theory.
The quantum corrections to the Higgs mass from the SM gauge boson and fermion loops generate 
quadratic sensitivity to a momentum cutoff scale such as the Planck scale.
Therefore, excessive fine-tuning between a bare Higgs mass term and the large quantum corrections 
would be needed to have the Higgs mass of electroweak (EW) scale, as observed at the LHC.

An alternative view of the naturalness problem for the SM was presented by Bardeen \cite{Bardeen:1995kv}. 
At the classical level, the SM Lagrangian is scale invariant if the Higgs mass term vanishes and
the Higgs mass term can be considered as a soft breaking parameter of the classical scale symmetry.
Still the Planck scale itself could not be explained in our theory since the quantum gravity is not included.
Instead our theory is required to be scale invariant at the Planck scale.  
Then the classical scale symmetry would protect the Higgs mass from the quadratic sensitivity up to the Planck scale.
The scale-invariant regularization for loop corrections on the Higgs mass automatically gets rid of quadratic terms of a momentum cutoff 
but just leaves logarithmic contributions,
and the theory is free from the fine-tuning in the Higgs mass renormalization.

The Higgs mass term itself can be generated by pure quantum effects in the classically scale-invariant theory.
Coleman and Weinberg (CW) introduced a mechanism of symmetry breaking, which relies on scale symmetry breaking 
by perturbative quantum loops \cite{Coleman:1973jx}.
They applied their mechanism to the scale-invariant SM, 
and their result implied that the Higgs boson mass should be around 10 GeV if there is no heavy fermions.
Now we know that it is not phenomenologically viable, because the Higgs mass was found to be about 125 GeV 
and the top quark turned out to be very heavy, 
which makes the radiatively induced Higgs mass negative from the CW mechanism of the EW symmetry breaking (EWSB).
Therefore, the SM should be extended in order to accommodate the CW mechanism for the viable Higgs mass.
The overall dominance of bosonic contributions to the Higgs effective potential over the fermion ones is required 
for the realization of the CW mechanism.

Many extensions of the SM for employing the CW mechanism have been suggested
\cite{Gildener:1976ih, Meissner:2006zh, Foot:2007as, Meissner:2007xv, Hambye:2007vf, Foot:2007iy,
	Iso:2009ss,	Iso:2009nw,	Holthausen:2009uc,	AlexanderNunneley:2010nw, Ishiwata:2011aa,
	Lee:2012jn, Okada:2012sg, Iso:2012jn, Gherghetta:2012gb, Das:2011wm,
	Carone:2013wla,	Khoze:2013oga, Farzinnia:2013pga, Gabrielli:2013hma, Antipin:2013exa,
	Hashimoto:2014ela, Hill:2014mqa, Guo:2014bha, Radovcic:2014rea, Binjonaid:2014oga,
	Davoudiasl:2014pya,	Allison:2014zya, Farzinnia:2014xia,	Pelaggi:2014wba, Farzinnia:2014yqa,
	Foot:2014ifa, Benic:2014aga, Guo:2015lxa, Oda:2015gna,	Das:2015nwk,
	Fuyuto:2015vna,	Endo:2015ifa, Endo:2015nba,	Plascencia:2015xwa,	Hashino:2015nxa,
	Karam:2015jta, Ahriche:2015loa,	Wang:2015cda, Haba:2015nwl,	Ghorbani:2015xvz,
	Helmboldt:2016mpi, Jinno:2016knw,	Ahriche:2016cio, Ahriche:2016ixu, Das:2016zue,
	Khoze:2016zfi, Karam:2016rsz, Oda:2017kwl, Ghorbani:2017lyk, Brdar:2018vjq,
	Brdar:2018num, YaserAyazi:2018lrv, YaserAyazi:2019caf,	Mohamadnejad:2019wqb, Jung:2019dog}.
Gildener and Weinberg (GW) extended the analysis of CW to theories which contain arbitrary numbers of scalar fields
and provided a formalism which allows a systematic minimization of the effective potential \cite{Gildener:1976ih}. 
In the GW formalism, we first minimize the tree-level potential and find a ``flat'' direction 
among the vacuum expectation values of the scalar fields at a particular renormalization scale. 
One of tree-level scalar masses turns out to be massless due to the flat direction. 
Including one-loop corrections in the potential, we give the potential a small curvature in the flat direction.
It generates the true physical vacuum, and the massless scalars become massive due to the radiative corrections.
%(We will adopt the GW formalism for the analysis of a classically scale invariant theory, which will be introduced in Section 2.)

Another pressing issue which requires an extension of the SM is 
the existence of nonbaryonic dark matter (DM) in the Universe,
since there is no proper DM candidate in the SM particle spectrum.
Although we have a lot of solid evidences for the DM from its gravitational interactions, 
its particle property is still in a mystery.
Among all the DM candidates, 
weakly interacting massive particle (WIMP) is the most popular one 
because of its natural mass and interaction ranges to give the right amount of the DM relic density observed today.
As a possible DM scenario, 
a singlet fermionic DM model was introduced in Refs.~\cite{Kim:2006af,Kim:2008pp}, 
which consists of a SM gauge singlet fermion and a singlet scalar in addition to the SM particles.
In that model, 
the singlet sector and the SM sector communicate with each other through Higgs portal interactions \cite{Patt:2006fw}, 
and the singlet fermion plays a role as a good WIMP candidate.
It has been shown that the singlet fermionic DM model is phenomenologically viable 
and provides interesting DM and collider phenomenology \cite{Kim:2009ke,Kim:2016csm,Kim:2018uov,Kim:2018ecv}. 

In this paper, we examine a singlet fermionic DM model which respects the classical scale symmetry. 
Particle masses in our model are obtained from the CW mechanism for the EWSB.
Since additional fermionic contributions to the effective potential come from the singlet fermions, 
besides the large top quark contribution, 
we need sufficient bosonic contributions 
in order to overcome the fermion ones and obtain viable Higgs and scalar masses.
For this purpose, we consider an additional $U(1)$ gauge symmetry in the singlet sector 
which introduces a dark gauge boson in addition to a dark scalar.
It turns out that this setup provides enough bosonic degrees of freedom 
to have a phenomenologically viable model for the CW mechanism, 
while simultaneously explaining the measured DM relic density
and satisfying the constraints from the DM direct detection experiments. 

This paper is organized as follows. 
In Sec. II, we describe the singlet fermionic DM model which has the Higgs portal interactions 
and respects the classical scale symmetry. 
Section III shows the effective potential for our model. 
The DM and collider phenomenology are discussed in Sec. IV and Sec. V, respectively.
Finally, Sec. VI is devoted to conclusions.

%%%%%%%%%%%%%%%%%%%%%%%%%%%%%%%%%%%%%%%%%%%%%%%%%%%%%%%%%%%%%%%%%
\section{model}
%%%%%%%%%%%%%%%%%%%%%%%%%%%%%%%%%%%%%%%%%%%%%%%%%%%%%%%%%%%%%%%%%

We consider a dark sector consisting of a classically massless complex scalar field $S$ 
and two Dirac fermion fields $\psi_i\ (i =1,2)$ which are the SM gauge singlets.  
The scalar mediator $S$ is responsible for the EWSB together with the SM Higgs doublet $H$, 
and the fermions $\psi_{1,2}$ are DM candidates,
with which the dark sector is gauged under $U(1)_X$ symmetry with new charge $X$. 
The extended Higgs sector Lagrangian with the renormalizable DM interactions is then given by
\beq \label{eq:DM_Lagrangian}
\mathscr{L}_{\rm DM} = \left(\mathcal{D}_\mu H\right)^{\dagger}\mathcal{D}^\mu H 
+ \left(D_{\mu} S\right)^\ast\left(D^{\mu} S\right) 
+ i\sum_{i=1,2}\left(\overline{\psi}_{iL}D\!\!\!\!/\psi_{iL} +\overline{\psi}_{iR}D\!\!\!\!/\psi_{iR}\right) 
- V_{S}(H,S) - V_F(\psi_{1,2},S), 
\eeq 
with the scale-invariant Higgs portal potential
\beq \label{eq:VS_potential}
V_S(H,S) =  \lambda_h (H^{\dagger} H)^2 + \lambda_{hs} H^{\dagger} H |S|^2 + \lambda_s |S|^4 ,
\eeq
and with the DM Yukawa interaction 
\beq \label{eq:VF_potential}
V_F(\psi,S) = g_{1S}\overline{\psi}_{1L}\psi_{1R} S + g_{2S}\overline{\psi}_{2L}\psi_{2R} S^\ast + \textrm{H.c.},
\eeq
where $g_{iS}$ are the DM Yukawa couplings and we assume $g_{1S}=g_{2S}=g_{S}$ for simplicity
although those are not necessarily the same in general.
As such, those DM fermions have the same mass and the equal portion in the relic abundance of the Universe.
The covariant derivative $\mathcal{D}^\mu$ is the usual SM one. 	
The new covariant derivative in the dark sector is defined as 
$D^\mu \varphi = \left(\partial^\mu + \ai g_X A_X^\mu X_\varphi\right)\varphi$ 
where $g_X$ and $A_X$ are the new dark gauge coupling and boson, respectively,
and the new fields are given as $\varphi = S, \psi_{1L}, \psi_{1R}, \psi_{2L}, \psi_{2R}$ 
with the following charge assignment for gauge anomaly cancellation:
\beq \label{eq:X-charge}
X_S = 1,\quad  X_{\psi_{1L}} = X_{\psi_{2R}} = \half,\quad  X_{\psi_{1R}} = X_{\psi_{2L}} = -\half.
\eeq
Note that the singlet fermionic DM field $\psi_i$ couples only to the singlet scalar $S$, and
the interactions of the singlet sector to the SM sector arise only through the Higgs portal $H^{\dagger}H$.  
The above Lagrangian obeys a local $U(1)_X$ symmetry 
under which all the SM fields are even while the other new fields transform as
$\varphi \rightarrow e^{i \alpha(x) X_\varphi }\varphi$.
One can of course assign the new $X$ charges differently 
and adopt a different type of anomaly free sets of fermions as in Ref.~\cite{Ahmed:2017dbb} 
(although their model is not conformally invariant), 
where a single Dirac fermion turns into two Majorana mass eigenstates.
Alternatively, we adopt two Dirac mass eigenstates as the DM candidates.
For different choices of the $X$ charge assignment, however, our numerical results will not change much, 
and can be applicable to other scenarios as far as the DM fermions have the same mass. 
With two Dirac fermions, one can simply extend the model with different DM masses ($g_{1S} \neq g_{2S}$),
which could be considered as the conformal version of a multicomponent DM model
and be useful in resolving the small-scale structure issues in galaxy formation.
However, such a general case is beyond scope of this paper, and we leave it for future studies. 

After the EWSB, the SM Higgs and the singlet scalar field
develop nonzero vacuum expectation values (VEVs) 
($v_h$, $v_s$) 
and can be written as
\beq
H = \left( \begin{array}{c} w^+ \\[1pt] \frac{1}{\sqrt{2}}\left(v_h + h +iw^0\right) \end{array} \right), \qquad
S = \frac{1}{\sqrt{2}}(v_s+s+i\chi),
\eeq
where $s$ and $\chi$ are CP-even and CP-odd states, respectively. 
Thus, the above Lagrangian is CP invariant.
Also, from the dark sector kinetic terms and the DM Yukawa interactions,
the $U(1)_X$ gauge boson and the DM fermions obtain their masses as
\beq
M_{A_X} = g_X v_s, \quad  M_\psi = g_S v_s/\sqrt{2}.
\eeq
Adopting the GW approach, we choose a flat direction among the scalar VEVs
along which the potential Eq.~(\ref{eq:VS_potential}) vanishes at some scale $\mu = \Lambda$. 
Along the flat direction, the potential minimization conditions
$\partial V/\partial H |_{\langle H^0\rangle=v_h/\sqrt{2}} = \partial V/\partial S |_{\langle S\rangle=v_s/\sqrt{2}}=0$
lead to the following relations
\beq \label{eq:lambdamin}
\lambda_{hs}(\Lambda) = - 2\lambda_h(\Lambda) /\tbt^2 , \qquad 
\lambda_s(\Lambda) = \lambda_h(\Lambda) / \tbt^4 ,
\eeq
where $\tbt\, (\equiv \tan\beta) = v_s/v_h$.
The neutral CP-even scalar fields $h$ and $s$ are mixed 
to yield the mass matrix given by
\beq \label{eq:mass_mattrix}
\mu_{h}^2 = 2 \lambda_h v_h^2 , \qquad
\mu_{s}^2 = 2 \lambda_h v_h^2 /\tbt^2 , \qquad
\mu_{hs}^2 = - 2 \lambda_h v_h^2 /\tbt.
\eeq
The corresponding scalar mass eigenstates $h_1$ and $h_2$ are admixtures of $h$ and $s$:
\beq
\left( \begin{array}{c} h_1 \\[1pt] h_2 \end{array} \right) =
\left( \begin{array}{cc} \cos \theta &\ -\sin \theta \\[1pt]
	 \sin \theta &\ \cos \theta \end{array} \right)
\left( \begin{array}{c} h \\[1pt] s \end{array} \right) ,
\eeq
where the mixing angle $\theta$ is given by
\beq \label{eq:tan_theta}
\tan \theta = \frac{y}{1+\sqrt{1+y^2}}, \qquad y \equiv  \frac{-2 \mu_{hs}^2}{\mu_h^2 - \mu_s^2}.
\eeq
Combining Eq.~(\ref{eq:mass_mattrix}) and Eq.~(\ref{eq:tan_theta}), 
we have $\tan\theta = -t_\beta$ or $1/t_\beta$.
The mixing angle $\tan\theta$ is expected to be very small 
(less than about 0.3 depending on the $h_2$ mass) due to 
the Large Electron-Positron (LEP) collider constraints \cite{Barate:2003sz}.
Through this scalar mixing $\theta$, 
there is a kinetic mixing between $U(1)_X$ and the SM $U(1)_Y$ gauge bosons arising in loop-level processes.
But it is proportional to the SM loops multiplied by the DM loops, so highly suppressed.	
If $\tan\theta = -t_\beta$, then $\lambda_s = \lambda_h/(\tan\theta)^4$ from Eq.~(\ref{eq:lambdamin})
so that $\lambda_s$ becomes very large.  
But this case is theoretically disfavored because of the perturbativity of the couplings.
Also, experimental constraints disfavor this scenario as well \cite{Farzinnia:2013pga, Farzinnia:2014yqa}.
Therefore, we only consider the case of $\tan\theta\, (\equiv \tth) = 1/t_\beta$,
which results in $\sin\theta\, (\equiv s_\theta) = c_\beta$ and $\cos\theta\, (\equiv c_\theta) = s_\beta$.
In this case, $\lambda_{hs}$ and $\lambda_s$ are suppressed by $\tth^2$ and $\tth^4$, respectively, 
which ensures the perturbativity of those couplings 
and induces the suppression of the Higgs portal interactions. 
As a result, the scalar couplings in Eq.~(\ref{eq:VS_potential}) change very slowly with $\Lambda$,
and similar discussions can be found also in Refs.~\cite{Gildener:1976ih, Hempfling:1996ht}.  

After diagonalizing the mass matrix, 
we obtain the physical masses of the two scalar bosons $h_1$ and $h_2$ as follows:
\beq \label{eq:scalar_mass}
M^2_1 = 2 \lambda_h v^2 \tth^2, \qquad M^2_2 = 0,
\eeq
where $v\ (\equiv \sqrt{v_h^2 + v_s^2})$ can be considered to be the VEV 
of the radial component of a scalar field composed of $h$ and $s$.
The value of $v$ is determined from the radiative corrections 
and is set to be the scale about $\Lambda$ according to GW. 
We assume that $M_1$ corresponds to the observed SM-like Higgs boson mass in what follows.
The SM Higgs $h_1$ has the tree-level mass while the new scalar singlet $h_2$ acquires its mass through radiative corrections, 
which is similar to the cases considered in Refs. \cite{Farzinnia:2014yqa, Ghorbani:2015xvz}.  
In terms of $h_1$ and $h_2$, the tree-level scalar potential in the flat direction can be expressed as
\beqr \label{eq:scalar_int}
 V_S(h_1, h_2) &=& \half M_1^2 h_1^2 +
 \frac{\lambda_h}{4}\Big[ (1-\tth^2)^2 h_1^4 + 4\tth(1-\tth^2)h_1^3(h_2+v) 
 + 4\tth^2 h_1^2 (h_2^2 + 2v h_2)  \nonumber \\[1pt]
&&  - 2\big\{ (1-\tth^2)h_1^2 + 2\tth h_1h_2 + 2v \tth h_1\big\}(\tilde{\chi}^2 - w^2) + (\tilde{\chi}^2 - w^2)^2 \Big],
\eeqr
where $\tilde{\chi} \equiv \chi/\tbt$ and $w^2 \equiv (w^0)^2 + 2w^+w^-$.
Note that we have discarded some of the scalar interaction terms in the potential in Eq.~(\ref{eq:scalar_int}) 
by imposing the constraints in Eq.~(\ref{eq:lambdamin}).
%For instance, the Higgs-scalar coupling $c_{122} \propto 1 - \tbt\tth$ for $h_1 h_2^2$ interaction vanishes
%due to our choice $\tth = 1/t_\beta$.
%Therefore, even if the radiatively generated $h_2$ mass is less than a half of the $h_1$ mass, 
%the partial decay width $\Gamma\left(h_1 \to h_2 h_2\right)$ is negligible 
%and our model is not constrained by the invisible Higgs decay measurements.

%%%%%%%%%%%%%%%%%%%%%%%%%%%%%%%%%%%%%%%%%%%%%%%%%%%%%%%%%%%%%%%%%
\section{Effective Potential}
%%%%%%%%%%%%%%%%%%%%%%%%%%%%%%%%%%%%%%%%%%%%%%%%%%%%%%%%%%%%%%%%%

The original approach of GW expressed the one-loop effective potential in terms of
the spherical coordinate (radial) field of the scalar gauge eigenstates.
Rather differently, we derive the effective potential with the physical eigenstates of the scalars
and obtain the scalar masses at one-loop level directly.
A similar study for a scalar DM is given in Ref.~\cite{Jung:2019dog}.
Let the background value of the physical scalar $h_i$ be $h_{ic}$. 
Then the effective potential is obtained by expanding the interaction terms in the Lagrangian 
around the background fields $h_{ic}$
and by keeping terms quadratic in fluctuating fields only.
From Eqs.~(\ref{eq:VF_potential}) and (\ref{eq:scalar_int}), the effective potential at one-loop level is given by
\beq
\Veff(h_{1c},h_{2c}) = V^{(0)}(h_{1c},h_{2c}) + V^{(1)}(h_{1c},h_{2c}),
\eeq
with
\beqr
V^{(0)}(h_{1c},h_{2c}) &=& \frac{\lambda_h}{4}\left[\left(1-\tth^2\right)^2h_{1c}^4
+4\tth(1-\tth^2)h_{1c}^3h_{2c} + 4\tth^2h_{1c}^2h_{2c}^2 \right],  \nonumber \\[1pt]
V^{(1)}(h_{1c},h_{2c}) &=& 
\sum_P n_P \frac{\bar{m}_P^4(h_{ic})}{64\pi^2}\left(\ln\frac{\bar{m}_P^2(h_{ic})}{\mu^2} - c_P\right),
\eeqr
where $c_P = 3/2\ (5/6)$ for scalars and fermions (gauge bosons) in the $\msbar$ scheme and
$\mu$ is a renormalization scale. 
$\bar{m}_P$ is a field-dependent mass and
the summation is over the particle species of fluctuating fields $P = h_{1,2}, w^0, w^\pm, \chi, Z, W^\pm, t, A_X, \psi_{1,2}$
and their degrees of freedoms ($n_P$) are given as follows:
\beq
n_{h_1}=n_{h_2}=n_{w^0}=n_\chi=1,\quad n_{w^\pm}=2,\quad n_Z=n_{A_X}=3,\quad n_{W^\pm}=6,
\quad n_t=-12, \quad n_{\psi_i}=-4.
\eeq
Taking the flat direction of the VEVs, 
we minimize the effective potential at $h_{1c} = 0$ and $h_{2c} = v$, 
which corresponds to $h_{c} = v_h$ and $s_{c} = v_s$ 
in terms of the background values of the scalar gauge eigenstates. 
The field-dependent mass $\bar{m}_P(h_{ic})$ is proportional to $h_{ic}$, 
so that $\bar{m}_P(h_{1c})$ is irrelevant to our study 
because $\partial \bar{m}_P(h_{1c})/\partial h_{1c} |_{h_{1c}=0}=0$.
The relevant field-dependent masses for $h_{2c}$ are obtained as 
\beqr \label{eq:eff_mass}
\bar{m}_{h_1}^2(h_{2c}) &=& \frac{3}{4}\left[\lambda_h s_{2\theta}^2 + \lambda_s s_{2\theta}^2
	+ \frac{\lambda_{hs}}{2}\left(\frac{1}{3} + c_{4\theta}\right)\right] h_{2c}^2
	= 2\lambda_h\tth^2h_{2c}^2,  \quad %\nonumber \\[1pt]
\bar{m}_{h_2}^2(h_{2c}) = 3\left(\lambda_h\sth^4 + \lambda_s\cth^4 
	+ \frac{1}{4}\lambda_{hs}s_{2\theta}^2\right)h_{2c}^2 = 0, \nonumber \\[1pt]
\bar{m}_{w^0}^2(h_{2c}) &=& \bar{m}_{w^\pm}^2(h_{2c}) 
	= \left(\lambda_h\sth^2 + \half\lambda_{hs}\cth^2\right)h_{2c}^2 = 0, \quad
\bar{m}_{\chi}^2(h_{2c}) = \left(\lambda_s\cth^2 + \half\lambda_{hs}\sth^2\right)h_{2c}^2 = 0, \nonumber \\[1pt]	
\bar{m}_{Z}^2(h_{2c}) &=& \frac{1}{4}\left(g_2^2+g_1^2\right)\sth^2h_{2c}^2
	= M_Z^2\frac{h_{2c}^2}{v^2}, \quad
\bar{m}_{W^\pm}^2(h_{2c}) = \frac{1}{4}g_2^2\sth^2h_{2c}^2
	= M_W^2\frac{h_{2c}^2}{v^2},	\quad
\bar{m}_{t}^2(h_{2c}) = \frac{y_t^2}{2}\sth^2h_{2c}^2
= M_t^2\frac{h_{2c}^2}{v^2},	\nonumber \\[1pt]
\bar{m}_{A_X}^2(h_{2c}) &=& g_X^2\cth^2h_{2c}^2
	= M_{A_x}^2\frac{h_{2c}^2}{v^2}, \quad
\bar{m}_{\psi_i}^2(h_{2c}) = \frac{g_S^2}{2}\cth^2h_{2c}^2
	= M_\psi^2\frac{h_{2c}^2}{v^2},
\eeqr
where the second equalities in the right-hand sides of the equations are obtained
by imposing the constraints in Eq.~(\ref{eq:lambdamin}). 

The masses of the physical scalars $h_{1,2}$ can be directly obtained by taking the second-order derivatives
of the effective potential with respect to the classical background fields $h_{ic}$ as
\beqr \label{eq:radit_mass}
M_1^2 &=& \frac{\partial^2\Veff}{\partial h_{1c}^2}\Big{|}_{\substack{h_{1c}=0 \\ h_{2c}=v}} 
	= 2 \lambda_h v^2 \tth^2, 	\nonumber \\[1pt]
M_2^2 &=& \frac{\partial^2\Veff}{\partial h_{2c}^2}\Big{|}_{\substack{h_{1c}=0 \\ h_{2c}=v}} 
	= \frac{1}{8\pi^2v^2}\left(M_1^4 + 6M_W^4 + 3M_Z^4 -12M_t^4 + 3M_{A_X}^4 -8M_\psi^4\right).
\eeqr
Although we have employed the strategy somewhat differently from those of earlier studies following the GW approach 
in Refs. \cite{Farzinnia:2014yqa, Ghorbani:2015xvz},
the final result for the scalar masses are equivalent. 
In total, we have four independent model parameters relevant for DM phenomenology.  
The four model parameters $\lambda_h$, $v_s$, $g_X$, and $g_S$  
determine the masses $M_{1,2}$, $M_{A_X}$, $M_\psi$, and the mixing angle $\theta$.
 The dependency of the model parameters are 
\beq \label{eq:parameters}
v = \frac{v_h}{\sth}, \quad v_s = \frac{v_h}{\tth}, \quad \lambda_h = \frac{M_1^2\cth^2}{2v_h^2}, 
\quad \lambda_{hs} = -\frac{M_1^2\sth^2}{v_h^2}, \quad \lambda_{s} = \frac{M_1^2\sth^2\tth^2}{2v_h^2},
\quad g_X = \frac{M_{A_X}\tth}{v_h}, \quad g_S = \sqrt{2}\frac{M_\psi \tth}{v_h} .
\eeq
Given the fixed Higgs mass $M_1$ and $v_h \simeq 246$ GeV, 
we constrain three independent new physics (NP) parameters by taking into
account various theoretical considerations and experimental measurements in the next section.

%%%%%%%%%%%%%%%%%%%%%%%%%%%%%%%%%%%%%%%%%%%%%%%%%%%%%%%%%%%%%%%%%
\section{Dark Matter Phenomenology}

 At present, the most accurate determination of the DM mass density $\Omega_{\rm DM}$ 
comes from global fits of cosmological parameters to a variety of observations
such as measurements of the anisotropy of the cosmic microwave background (CMB) data by the Planck experiment 
and of the spatial distribution of galaxies \cite{Tanabashi:2018oca}:
\beq
\Omega_{\rm CDM}h^2 = 0.1186 \pm 0.0020.
\label{eq:relic_obserb}
\eeq
This relic density observation will exclude some regions in the model parameter space.
The relic density analysis in this section includes all possible channels of
$\psi_i\psi_i$ pair annihilation into the SM particles.  	
In this work, we implement the model described in Sec. II into the CalcHEP package \cite{Belyaev:2012qa}. 
Using the numerical package micrOMEGAs \cite{Belanger:2018mqt} 
that utilizes the CalcHEP for computing the relevant annihilation cross sections, 
we compute the DM relic density and the spin-independent DM-nucleon scattering cross sections.
Especially, micrOMEGAs is known to be effective for the relativistic treatment of the thermally averaged cross section
and for a precise computation of the relic density in the region where annihilation 
through a Higgs exchange occurs near resonance \cite{Belanger:2004yn}.

%%%%%%%%%%%%%%%%%%%%%%%%%%%%%%%%%%%%%%%%%%%%%%%%%%
\begin{figure}[!hbt]
	\centering%
	\subfigure[ ]{\label{omegahr1a} %
		\includegraphics[width=7.6cm]{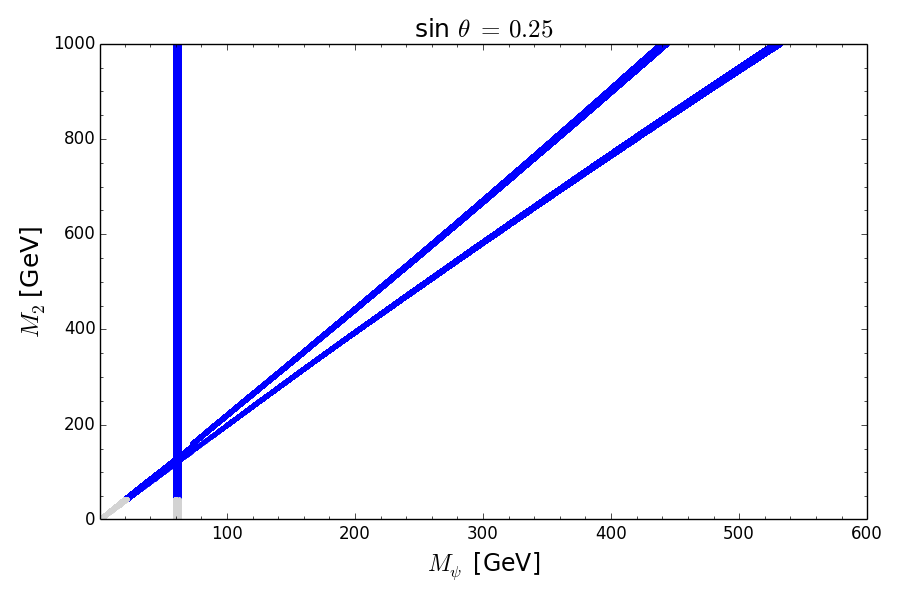}} \
	\subfigure[ ]{\label{omegahr1b} %
		\includegraphics[width=7.6cm]{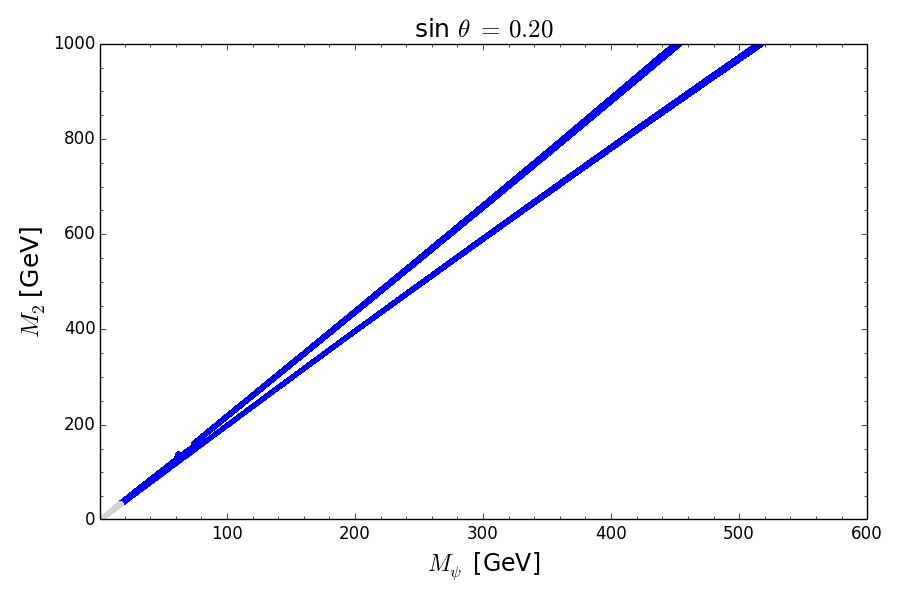}} \\   
	\subfigure[ ]{\label{omegahr2a} %
		\includegraphics[width=7.6cm]{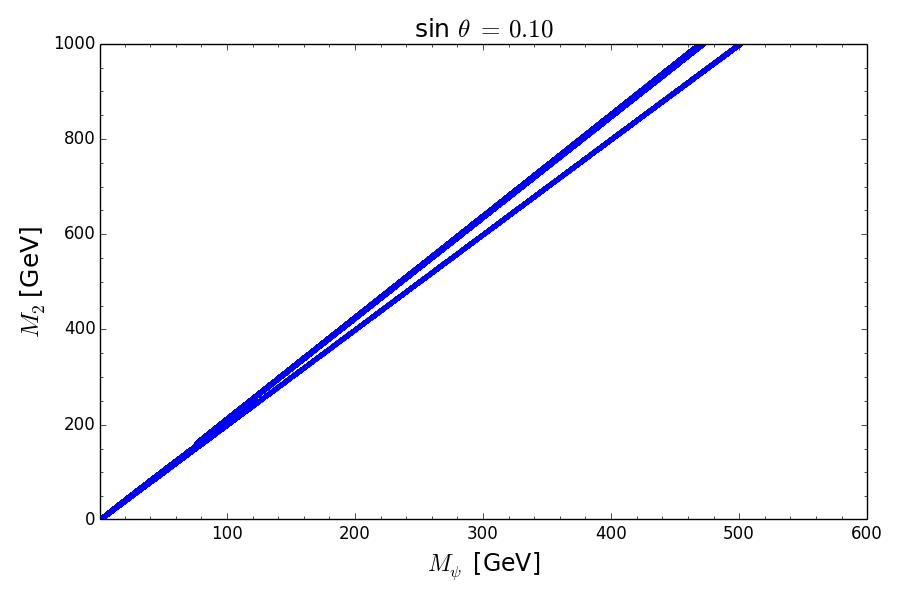}} \
	\subfigure[ ]{\label{omegahr2b} %
		\includegraphics[width=7.6cm]{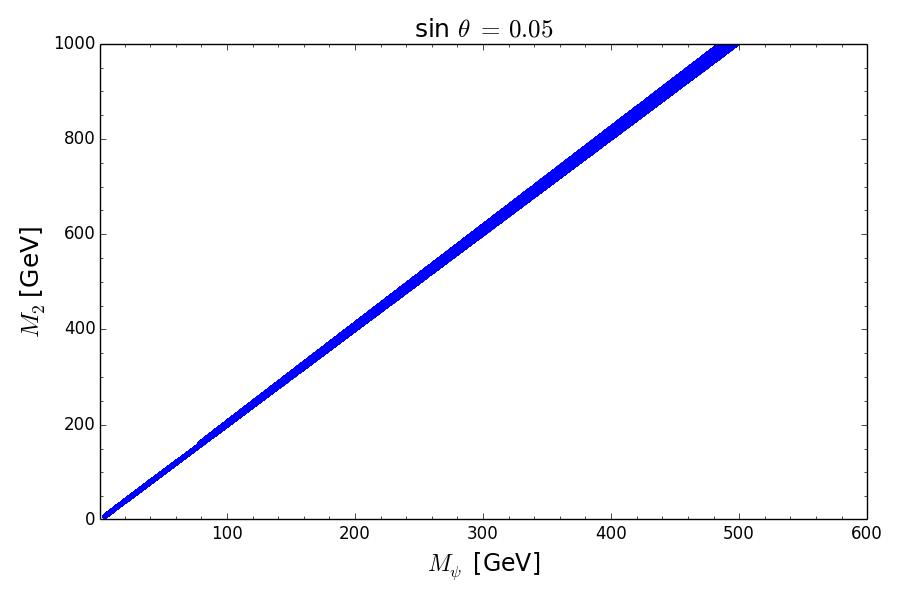}}     
	\caption{Allowed regions for the parameter sets of ($M_\psi, M_2$) by relic density observations at $3\sigma$ level
		for $\sin\theta = 0.25, 0.2, 0.1, 0.05$. 
	    For $\sin\theta =$ 0.25 and 0.2, some of allowed parameter sets by the relic density observation in low DM
	    mass (grayed) region are excluded by the LEP constraint. } 
	\label{fig:relicdensity}
\end{figure}
%%%%%%%%%%%%%%%%%%%%%%%%%%%%%%%%%%%%%%%%%%%%%%%%%%

%%%%%%%%%%%%%%%%%%%%%%%%%%%%%%%%%%%%%%%%%%%%%%%%%%
\begin{figure}[!hbt]
	\centering%
	\includegraphics[width=8cm]{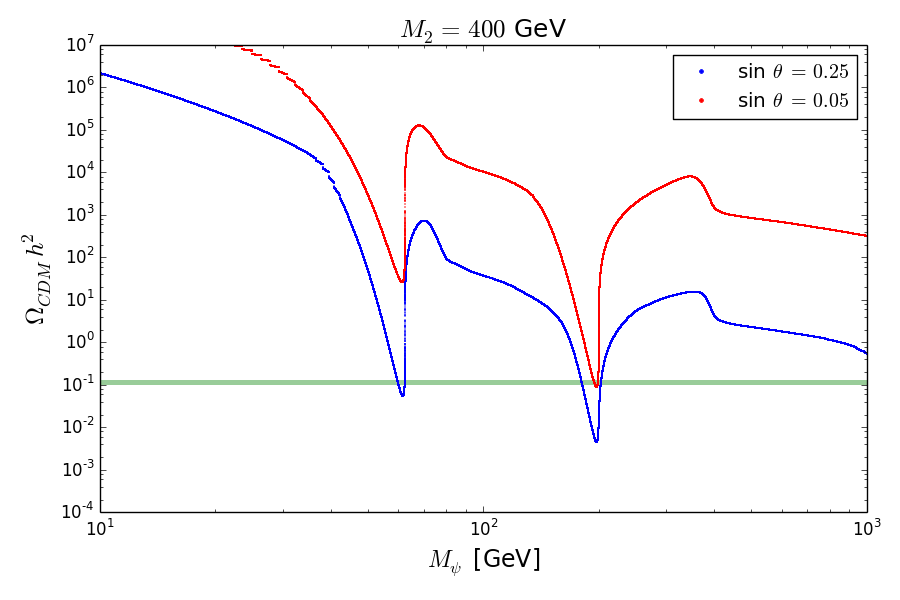}
	\caption{Relic density $\Omega_{\rm CDM}h^2$ as a function of $M_\psi$ 
		for $\sin\theta = 0.25, 0.05$, and $M_2 = 400$ GeV.
		The shaded (green) parallel band is allowed by the recent observation given in Eq.~(\ref{eq:relic_obserb}).
	}
	\label{fig:relic_func}
\end{figure}
%%%%%%%%%%%%%%%%%%%%%%%%%%%%%%%%%%%%%%%%%%%%%%%%%%

For an illustration of allowed new model parameter spaces,
we choose the benchmark points for the scalar mixing angle $\sin\theta = 0.25, 0.2, 0.1, 0.05$
and perform the numerical analysis by varying the following two NP parameters: $M_\psi$, $M_2$.
The new gauge boson mass $M_{A_X}$ is determined by $M_\psi$ and $M_2$ from Eq.~(\ref{eq:radit_mass}),
and the dependency of other NP parameters are shown in Eq.~(\ref{eq:parameters}).
In order to see the relic density constraints on the singlet fermionic DM interaction,
we first plot the allowed region of the DM mass $M_\psi$ and the scalar mediator mass $M_2$ 
constrained by the current relic density observations at 3$\sigma$ level for four different values of $\sin\theta$
in Fig.~\ref{fig:relicdensity}.
Due to the small mixing $\theta$ suppression of the Higgs portal couplings $\lambda_{hs}$ and $\lambda_s$,
the allowed regions by the relic density observation only appear near the resonance regions 
of $h_1$ and $h_2$ masses.
Especially near the $h_2$ resonance region, 
the central areas inside the allowed regions in the figures are cut off by the stringent relic density constraint.   
This behavior can be clearly understood in Fig.~\ref{fig:relic_func}
which plots the relic density as a function of $M_\psi$ for $\sin\theta = 0.25, 0.05$, and $M_2 = 400$ GeV.
The allowed region by the relic observation constraint is shown as the green parallel band in the figure, 
and it passes through the $h_1$ and $h_2$ mass resonance region for $\sin\theta = 0.25$ case
but barely touches the $h_2$ mass resonance region only for $\sin\theta = 0.05$.  
We found that $\sin\theta \gtrsim 0.05$ is the lower bound to explain the current relic observation in this model. 

%%%%%%%%%%%%%%%%%%%%%%%%%%%%%%%%%%%%%%%%%%%%%%%%%%
\begin{figure}[!hbt]
	\centering%
	\subfigure[ ]{\label{sigmaSIr1a} %
		\includegraphics[width=7.6cm]{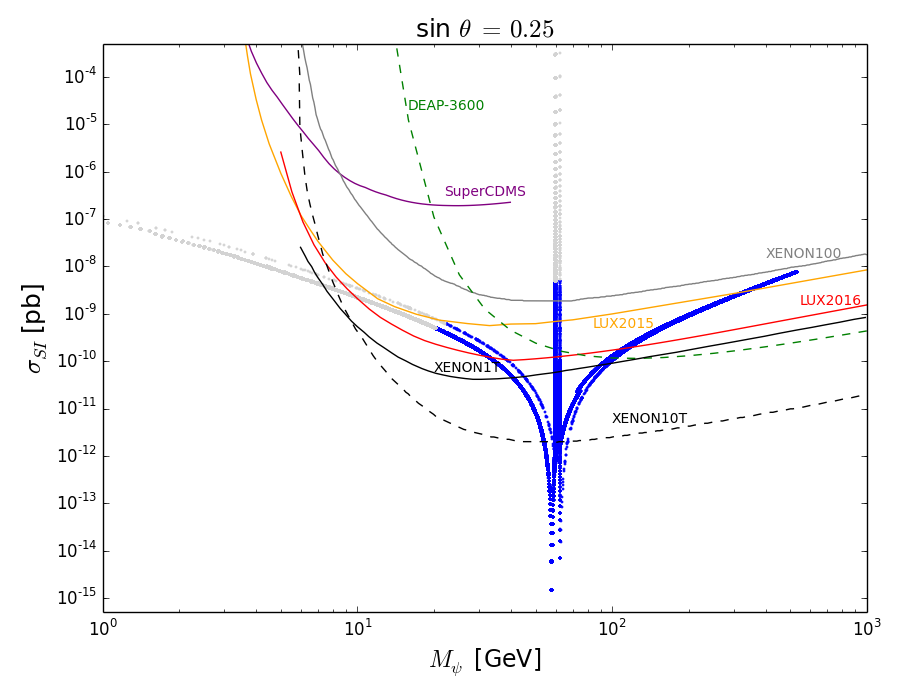}}  \
	\subfigure[ ]{\label{sigmaSIr1b} %
		\includegraphics[width=7.6cm]{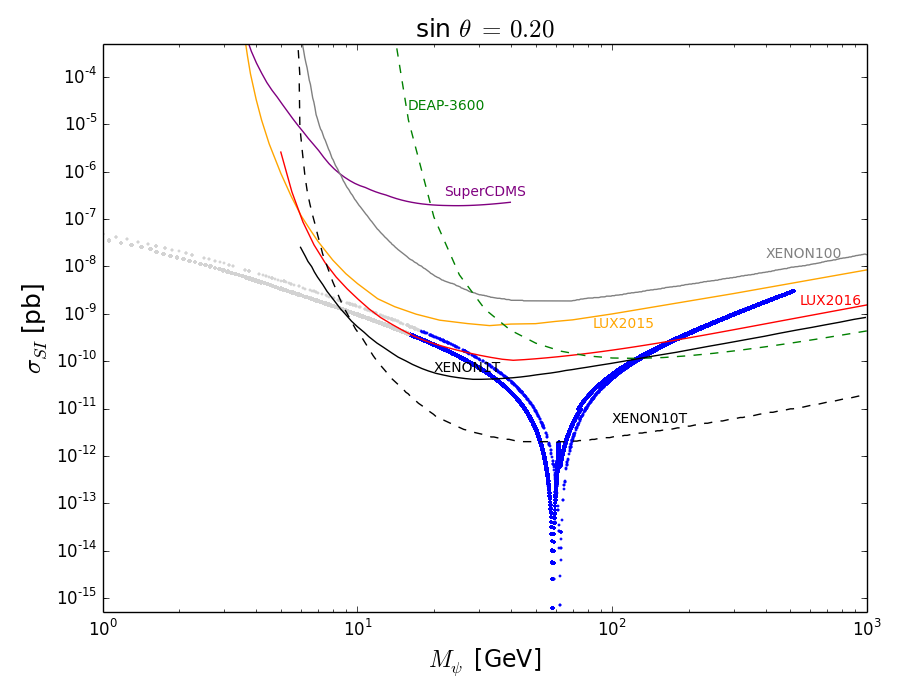}}   \\
	\subfigure[ ]{\label{sigmaSIr2a} %
		\includegraphics[width=7.6cm]{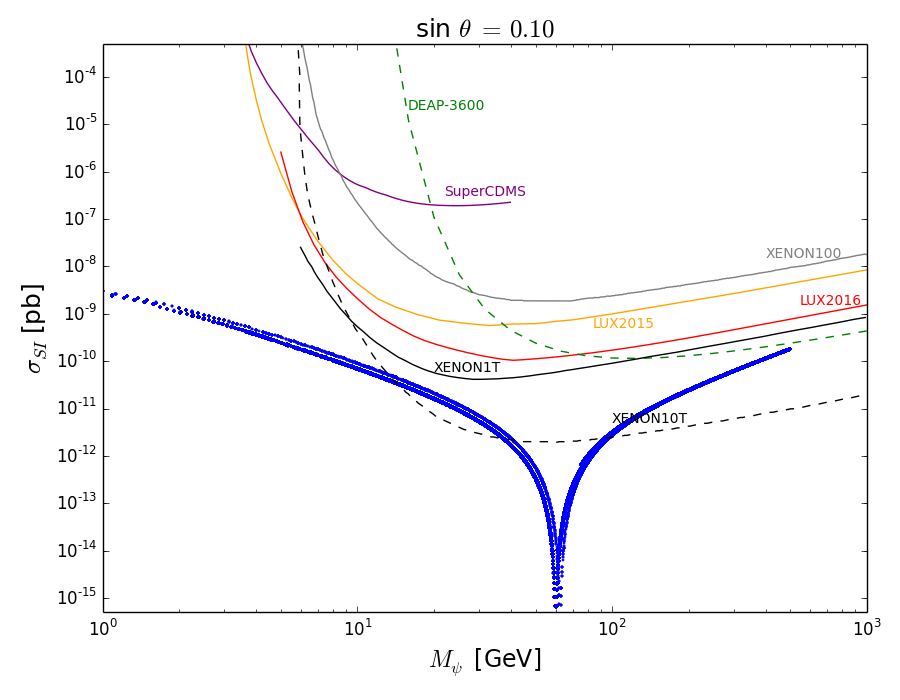}}  \
	\subfigure[ ]{\label{sigmaSIr2b} %
		\includegraphics[width=7.6cm]{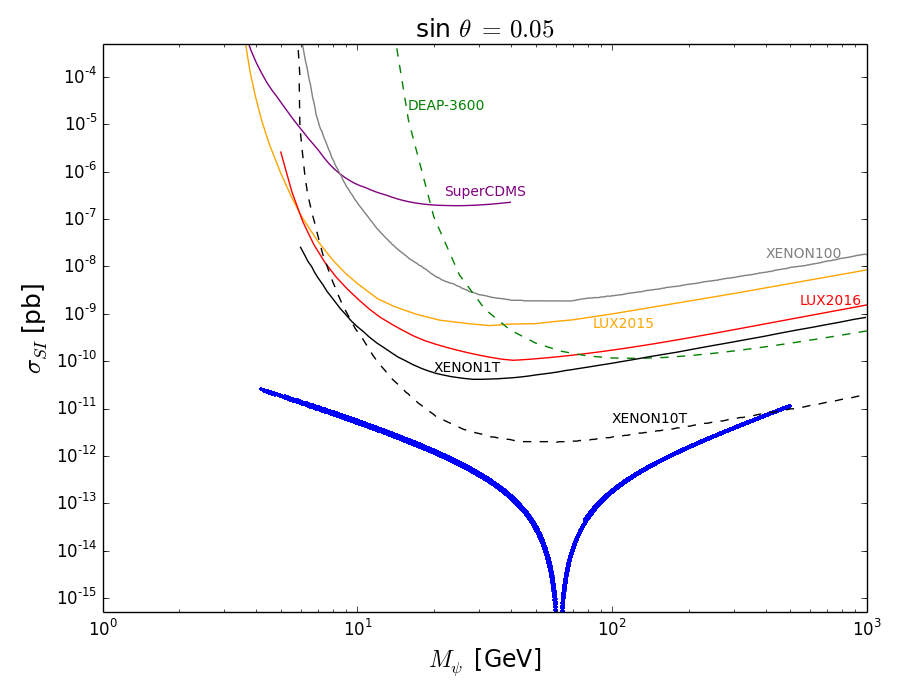}}   
	\caption{Spin-independent DM-nucleon scattering results allowed by relic density observations.  
		Also shown are the observed limits from LUX 2015, 2016, XENON100, XENON1T, SuperCDMS,   
		and expected limits from DEAP-3600 and XENON10T. 
	    For $\sin\theta =$ 0.25 and 0.2, some of allowed parameter sets by the relic density observation in low DM
	    mass (grayed) region are excluded by the LEP constraint.}
	\label{fig:sigmaSI}
\end{figure}
%%%%%%%%%%%%%%%%%%%%%%%%%%%%%%%%%%%%%%%%%%%%%%%%%%

In Fig.~\ref{fig:sigmaSI}, 
we plot the spin-independent DM-nucleon scattering cross section by varying the DM mass $M_\psi$ 
with parameter sets allowed by the relic density observation and compare the results with the observed upper limits
obtained at 90\% level from LUX 2015, 2016, XENON100, XENON1T, SuperCDMS, 
and with the expected limits from DEAP-3600 and XENON10T \cite{Akerib:2016vxi, Aprile:2018dbl}.
For $\sin\theta =$ 0.25 and 0.2, some of allowed parameter sets by the relic density observation in low DM mass (grayed) region 
are excluded by the LEP constraint also as in Fig.~\ref{fig:relicdensity},
and parameter sets only near $h_1$ resonance region survive.
For $\sin\theta =$ 0.1 and 0.05, most of the allowed regions by the relic constraints are not excluded 
by these direct detection bounds.
We obtained the approximate upper bound, $\sin\theta < 0.28$, from the XENON1T and LEP2 constraints. 
The DM-nucleon scattering cross section is dominated by the tree-level t-channel scalar exchange processes,
and there is a destructive interference between $H_1$ and $H_2$ contributions
 to the scattering amplitude due to orthogonality of the Higgs mixing matrix.
This behavior is a very generic aspect in case there are extra singlet scalar bosons mixing with the SM Higgs boson \cite{Kim:2006af}. 	
If $h_2$ mass is close to $h_1$ mass, 
the cancellation between the t-channel $h_1$ and $h_2$ exchange contributions becomes significant,
and as a result the scattering cross section is drastically reduced in the region near $M_\psi \sim M_1/2$ 
as shown in the figure.
Since $h_2$ mass is obtained by the radiative corrections, 
its mass scale is expected to be similar to the electroweak scale and also to the GW scale satisfying Eq.~(\ref{eq:lambdamin}).
Also, if one adopts the conditions that lead to a strong first order phase transition as needed to produce
the observed baryon asymmetry of the Universe, $h_2$ mass should not exceed 1 TeV \cite{Profumo:2007wc}.
Therefore, we do not expect it to be too large and so assume it smaller than 1 TeV.
Due to the relic constraint, the value of $M_\psi$ is only allowed near the resonance region of the scalar masses
(especially $h_2$ mass), 
and the DM mass heavier than about 530 GeV is disfavored as shown in Figs.~\ref{fig:relicdensity} and \ref{fig:sigmaSI}.
Note also that small DM mass is disfavored as well if $\sin\theta$ is too small (or large),
and a few GeV of DM mass is allowed only when $\sin\theta$ is about 0.1.
%For $\sin\theta \geq 0.2$, the DM mass heavier than about 130 GeV is disfavored.
For $\sin\theta =$ 0.2 (0.25), the DM mass heavier than about 130 (100) GeV is constrained 
by the XENON1T bound.

%%%%%%%%%%%%%%%%%%%%%%%%%%%%%%%%%%%%%%%%%%%%%%%%%%%%%%%%%%%%%%%%%
\section{Collider Phenomenology}

%%%%%%%%%%%%%%%%%%%%%%%%%%%%%%%%%%%%%%%%%%%%%%%%%%
\begin{figure}[!hbt]
	\centering%
	\includegraphics[width=8cm]{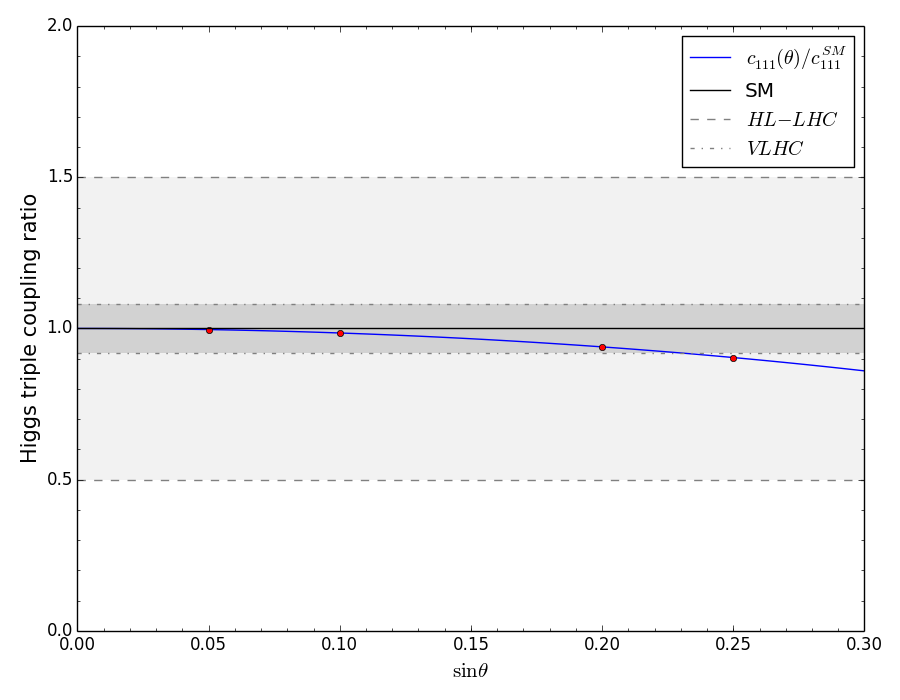}
	\caption{Higgs triple coupling ratio $c_{111}/c_{111}^\textrm{SM}$ as a function of $\sin\theta$.
		Dashed and dot-dashed lines represent the expected experimental precisions on $c_{111}$
		at future hadron colliders, HL-LHC and VLHC, respectively.  }
	\label{fig:Branchingf}
\end{figure}
%%%%%%%%%%%%%%%%%%%%%%%%%%%%%%%%%%%%%%%%%%%%%%%%%%

In this section, we discuss phenomenological implications and interactions of new particles 
in our model at colliders.
In general, Higgs interactions can be significantly modified
due to the Higgs portal terms given in Eq.~(\ref{eq:VS_potential}) in this model.
For instance,
the triple Higgs coupling $c_{111}$ for $h_1^3$ interaction is important for the collider phenomenology 
and will be probed through Higgs pair production at future colliders \cite{Dawson:2013bba}.
We read out $c_{111}$ from Eq.~(\ref{eq:scalar_int}) as
\beq
c_{111} = 6\lambda_h v \tth(1 - \tth^2) =\ \cth(1 - \tth^2)c_{111}^\textrm{SM}, 
\eeq
where the SM value $c_{111}^\textrm{SM} = 3 M_1^2/v_h$, 
and the ratio of the triple Higgs coupling $c_{111}/c_{111}^\textrm{SM}$
depends on the mixing angle as shown in Fig.~\ref{fig:Branchingf}.
%One can see that the $c_{111}$ reduces to the SM value when $\theta = 0$.
It shows that it would be very difficult to observe the deviation in the HL-LHC experiment 
but marginally possible to see it in the VLHC experiment.
%We may observe the deviation in the High Luminosity LHC era.

Now we consider the SM Higgs boson decays into the new particles.
%Through the triple couplings 
The SM Higgs boson $h_1$ and new scalar $h_2$ 
can decay into one another depending on their masses.
If $M_2 \leq M_1/2$, it is kinematically allowed that
$h_1$ decays into a pair of $h_2$ so that the total decay width of $h_1$ increases. 
%with the partial decay width,
%\beq
%\Gamma\left(h_1 \to h_2 h_2\right)
%= \frac{\lambda_h^2v_h^2\sqrt{M_1^2-4M_2^2}}{2\pi M_1^2}
%                    \left(1 - \tbt\tth\right)^2\sth^2\tth^2.
%\eeq
However, the $h_1h_2^2$ interaction is absent in our case because the relevant Higgs-scalar coupling 
$c_{122}\propto 1-t_\beta t_\theta$ vanishes under the constraint $t_\theta=1/t_\beta$.
Therefore, the partial decay width $\Gamma\left(h_1 \to h_2 h_2\right)$ is negligible.

%%%%%%%%%%%%%%%%%%%%%%%%%%%%%%%%%%%%%%%%%%%%%%%%%%
\begin{figure}[!hbt]
	\centering%
	\includegraphics[width=8cm]{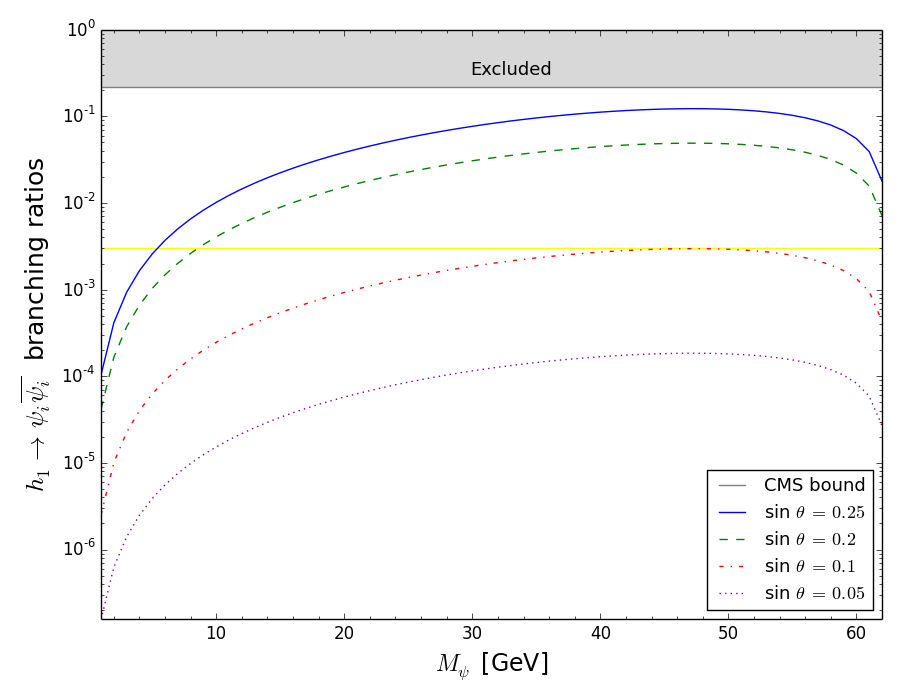}
	\caption{Branching fractions of 
		$h_1 \rightarrow \psi\overline{\psi}$ decays are shown
		as different lines for four values of
		$\sin\theta$ = 0.25 (solid blue), 0.2 (dashed green), 
		0.1 (dot-dashed red), 0.05 (dotted purple).
		The shaded region is excluded by the 95\% C.L. limit of BR($h_1 \to$ inv.) $< 0.22$,
		and the (yellow) parallel line represents the expected experimental precision at ILC. }
	\label{fig:invhiggs}
\end{figure}
%%%%%%%%%%%%%%%%%%%%%%%%%%%%%%%%%%%%%%%%%%%%%%%%%%

If $M_\psi < M_1/2$, still, 
the SM Higgs boson $h_1$ can decay into a pair of the DM with decay width,
\beq
\Gamma_\textrm{inv.} = \Gamma\left(h_1 \to \psi_i\bar{\psi_i}\right)
= \frac{g_S^2M_1 \sth^2}{8\pi}\left(1 - \frac{4M_\psi^2}{M_1^2}\right)^{3/2}
= \frac{M_1 M_\psi^2 \sth^2 \tth^2}{4\pi v_h^2}\left(1 - \frac{4M_\psi^2}{M_1^2}\right)^{3/2},
\eeq 
where the final states are summed over $i = 1,2$ 
and the width contributes to the branching fraction of the invisible Higgs decays,
BR($h_1 \to$ inv.) = $\Gamma_\textrm{inv.}
/(\Gamma_\textrm{SM} + \Gamma_\textrm{inv.})$,
where $\Gamma_\textrm{SM}$ = 4.07 MeV \cite{Heinemeyer:2013tqa}.
We depict the modified invisible Higgs decay branching ratio in Fig.~\ref{fig:invhiggs}
by applying the combined 90\% C.L. limit of BR($h_1 \to$ inv.) $< 0.22$ to our model.
The most recent upper limit on the Higgs invisible decay has been set
by the CMS Collaboration combined with the run II data set with the luminosity of 35.9 fb$^{-1}$
at a center-of-mass energy of 13 TeV \cite{Khachatryan:2016whc, Sirunyan:2018owy}.
For future colliders, the International Linear Collider (ILC) would provide a upper limit BR($h_1 \to$ inv.) $< 0.003$ 
at 95$\%$ C.L. \cite{Barklow:2017suo, Fujii:2017vwa}.
One can clearly see from the figure that
our model is not constrained by the current invisible Higgs decay measurements	
but can be tested for $\sin\theta > 0.1$ at a future ILC experiment.

The singletlike Higgs $h_2$ production and decays into the SM particles at colliders 
are similar to those of the SM Higgs boson
except for the suppression factor $\sin^2 \theta$ (and ignoring the exotic channels such as $h_2 \rightarrow \psi\bar{\psi}$).
If  $h_2$ is light enough,
the LEP bound on the light Higgs boson should be applied \cite{Barate:2003sz}.
The LEP experiments provide a 95$\%$ C.L. upper bound on the mixing angle $\sin^2\theta$
as a function of $h_2$ mass,
which corresponds to $\sin^2\theta \simeq O(0.01)$ for $M_2 = 20$ GeV and
$\sin^2 \theta \simeq O(0.1)$ for $M_2 = 100$ GeV. 
This LEP bound was taken into account of the DM phenomenology in the previous section 
in Figs.~\ref{fig:relicdensity} and \ref{fig:sigmaSI}.

If the mass of $h_2$ is in the range of $M_1 < M_2 < 2 M_1$, 
it will decay dominantly to the pairs of $WW$ and $ZZ$.
If $M_2 > 2 M_1$, the $h_2 \to h_1 h_1$ channel also opens 
through the triple coupling,
\beq
c_{112}=4 \lambda_h v t_\theta^2 = 2 \frac{M_1^2}{v_h} s_\theta.
\eeq
Moreover, if kinematically allowed, $h_2$ may also decay into the DM pairs.
However, for the phenomenologically allowed parameter space, it turns out that those additional channels 
$h_2 \to h_1 h_1$ and $h_2 \to \psi_i \bar{\psi_i}$
are not so significant.
It implies that also in that mass range $h_2$ dominantly decays to a pair of SM gauge bosons.
Then, we obtain a strong constraint for the additional Higgs boson 
from the current LHC searches for a new scalar resonance 
decaying to a pair of $Z$ bosons \cite{Khachatryan:2015cwa, Aaboud:2017rel, Sirunyan:2018qlb}.
The 95 $\%$ C.L. upper bound on the production rate of scalar resonance $X$, 
$\sigma(pp\rightarrow X \rightarrow ZZ)$, at the LHC, can be translated to the upper bound
on the mixing angle $\sin^2\theta$ for a given $M_2$.
We obtain the upper bounds, for instance, $\sin^2\theta \lesssim 0.06$ (strongest) for $M_2 =$ 200 GeV, 
$\sin^2\theta \lesssim 0.09$ for $M_2 =$ 500 GeV and
$\sin^2\theta \lesssim 0.36$ for $M_2 =$ 1 TeV.
From this, we found $\sin\theta \lesssim 0.25$ is the marginally allowed upper bound
and used the value for our numerical analysis. 		

For future colliders, the bounds from the ILC experiment can significantly supersede the LEP bounds.
The 95$\%$ C.L. upper bounds on the mixing angle $\sin^2\theta$ are expected to be $O(0.001)$ for small $h_2$ mass 
below about 140 GeV \cite{Drechsel:2018mgd}.
On the other hand, for a larger $h_2$ mass, the current LHC bounds will be improved by an order of magnitude 
at the HL-LHC with an integrated luminosity of 3000 fb$^{-1}$.

Besides $h_2$ and $\psi_i$,
one may expect to see the collider signatures of the new gauge boson $A_X$ at future colliders.
The mass of $A_X$ is constrained by the mass relation in Eq. (16) and generically large, $M_{A_X} > 240$ GeV. 
Thus, in  most of the region of parameter space where we have interest in,
$M_{A_X} > 2 M_\psi$ so that $A_X$ can decay into a pair of DM with decay width,
\beq
\Gamma\left(A_X \to \psi_i\bar{\psi_i}\right)
= \frac{g_X^2M_{A_X}}{24\pi}
\left(1 - \frac{4M_\psi^2}{M_{A_X}^2}\right)^{3/2}
= \frac{M_{A_X}^3\tth^2}{24\pi v_h^2}
\left(1 - \frac{4M_\psi^2}{M_{A_X}^2}\right)^{3/2},
\eeq 
where the final states are summed over $i = 1,2$.
On the other hand, $A_X$ might be produced at colliders in a pair
through the off-shell $h_{1,2}$ decays
while not from the on-shell $h_2$ decays
because $A_X$ is much heavier than $h_2$.
If $A_X$ is produced from off-shell $h_{1,2}$ decays,
these are suppressed by the small mixing angle and kinematics.
Therefore, it is hard to investigate the collider phenomenology of $A_X$
at the LHC and the near future colliders.

\section{Concluding Remarks}

In this work, we investigated an extension of the SM which is renormalizable and classically scale invariant.
We introduced the SM gauge singlet DM sector that consists of a complex scalar field $S$, 
a $U(1)$ gauge boson field $A_X$, and an anomaly free set of two Dirac fermion DM fields $\psi_{1,2}$.
The communication between the SM and the singlet DM sectors is accomplished by the Higgs portal interaction.
The scalar masses are generated quantum mechanically through the Coleman-Weinberg mechanism 
for the EWSB.
Given the SM-like Higgs mass $M_1 \simeq 125$ GeV and the VEV $v_h \simeq 246$ GeV, 
we have three independent new physics parameters.
We chose the Higgs mixing angle $\sin\theta$, the singlet scalar mass $M_2$, 
and the singlet fermionic DM mass $M_\psi$ as the three free model parameters
for further phenomenological studies,
letting the new gauge boson mass $M_{A_X}$ be determined by  $M_2$ and $M_\psi$.

In the early Universe, the singlet fermionic dark matters are pair annihilated into SM particles, 
mainly through s-channel scalar mediations.
Since the scalar mixing angle $\sin\theta$ should be small 
due to the current experimental constraints on Higgs boson,
we need scalar resonance effects to acquire a large enough annihilation cross section for the DM pair annihilation.
Therefore, the right amount of DM relic density is obtained when the scalar masses are about twice the DM mass, 
i.e, $M_1 \simeq 2 M_\psi$ or $M_2 \simeq 2 M_\psi$, as was shown in Fig.~\ref{fig:relicdensity}. 

For the direct detection of the DM, 
we obtained a small enough cross section for the spin-independent elastic scattering of the DM with a nuclei,
through the small scalar mixing angle $\sin\theta$ and 
due to the cancellation between the t-channel $h_1$ and $h_2$ exchange diagrams, 
which occurs significantly if $M_1 \simeq M_2$.
Therefore, our model is phenomenologically viable, 
which has new model parameter sets satisfying all of the current experimental constraints. 
Taking into account the LEP and LHC constraints, the viable range of the scalar mixing angle lies
between $0.05 \leq \sin\theta \leq 0.25$, 
and the allowed range of the DM mass is highly constrained by that angle 
due to the relic and the direct detection constraints.

Our model may also reveal observable signatures at colliders, so that it can be tested further.
For some model parameter ranges, a small deviation of the triple Higgs coupling might be observed at the VLHC, and
the invisible Higgs decay would be detected at a future lepton collider such as the ILC.
More direct probes of the present models come from new particle searches.
The new scalar particle $h_2$ would be detected through the $e^+e^- \rightarrow Z h_2$ process at future lepton colliders,
or through $pp \rightarrow h_2 \rightarrow ZZ\, (WW)$ processes at future hadron colliders, depending on the mass and couplings.
Therefore, the obtained allowed parameter sets in this model can be used as benchmark points 
to test proper DM model candidates as the future experimental progress can further improve the bounds.

\acknowledgments
This work was supported by the Basic Science Research Program
through the National Research Foundation of Korea (NRF)
funded by the Ministry of Science, ICT, and Future Planning under the 
Grant No. NRF-2017R1E1A1A01074699 (S.-h. Nam), 
and funded by the Ministry of Education under the Grants No. NRF-2016R1A6A3A11932830 (S.-h. Nam), 
No. NRF-2018R1D1A3B07050649 (K.~Y. Lee), and No. NRF-2018R1D1A1B07050701 (Y.~G. Kim).

%%%%%%%%%%%%%%%%%%%%%%%%%%%%%%%%%%%%%%%%%%%%%%%%%%
\def\npb#1#2#3 {Nucl. Phys. B {\bf#1}, #2 (#3)}
\def\plb#1#2#3 {Phys. Lett. B {\bf#1}, #2 (#3)}
\def\prd#1#2#3 {Phys. Rev. D {\bf#1}, #2 (#3)}
\def\jhep#1#2#3 {J. High Energy Phys. {\bf#1}, #2 (#3)}
\def\jpg#1#2#3 {J. Phys. G {\bf#1}, #2 (#3)}
\def\epj#1#2#3 {Eur. Phys. J. C {\bf#1}, #2 (#3)}
\def\arnps#1#2#3 {Ann. Rev. Nucl. Part. Sci. {\bf#1}, #2 (#3)}
\def\ibid#1#2#3 {{\it ibid.} {\bf#1}, #2 (#3)}
\def\none#1#2#3 {{\bf#1}, #2 (#3)}
\def\mpla#1#2#3 {Mod. Phys. Lett. A {\bf#1}, #2 (#3)}
\def\pr#1#2#3 {Phys. Rep. {\bf#1}, #2 (#3)}
\def\prl#1#2#3 {Phys. Rev. Lett. {\bf#1}, #2 (#3)}
\def\ptp#1#2#3 {Prog. Theor. Phys. {\bf#1}, #2 (#3)}
\def\rmp#1#2#3 {Rev. Mod. Phys. {\bf#1}, #2 (#3)}
\def\zpc#1#2#3 {Z. Phys. C {\bf#1}, #2 (#3)}
\def\cpc#1#2#3 {Chin. Phys. C {\bf#1}, #2 (#3)}
%%%%%%%%%%%%%%%%%%%%%%%%%%%%%%%%%%%%%%%%%%%%%%%%%%

\end{document}